\documentclass[prd,twocolumn,nofootinbib]{revtex4}

\usepackage{graphicx}
\usepackage{amsmath,amssymb,amsfonts}
\newcommand{\be}{\begin{equation}}
\newcommand{\ee}{\end{equation}}
\newcommand{\ba}{\begin{eqnarray}}
\newcommand{\ea}{\end{eqnarray}}
\newcommand{\ban}{\begin{eqnarray*}}
\newcommand{\ean}{\end{eqnarray*}}

\newcommand \pp {$pp$}

\begin{document}

\title{A New Phase of Matter: Quark-Gluon Plasma\\ Beyond the Hagedorn Critical Temperature}
\author{Berndt M\"uller}\thanks{To appear in {\em Melting Hadrons, Boiling Quarks} by Rolf Hagedorn and Johan Rafelski (editor), Springer Publishers, 2015}
\affiliation{Department of Physics, Duke University, Durham, NC 27708-0305, USA}
\affiliation{Brookhaven National Laboratory, Upton, NY 11973, USA}

\begin{abstract} 
I retrace the developments from Hagedorn's concept of a limiting temperature for hadronic matter to the discovery and characterization of the quark-gluon plasma as a new state of matter. My recollections begin with the transformation more than 30  years ago of Hagedorn's original concept into its modern interpretation as the ``critical'' temperature separating the hadron gas and quark-gluon plasma phases of strongly interacting matter. This was followed by the realization that the QCD phase transformation could be studied experimentally in high-energy nuclear collisions. I describe here my personal effort to help develop the strangeness experimental signatures of quark and gluon deconfinement and recall how the experimental program proceeded soon to investigate this idea, at first at the SPS, then at RHIC, and finally at LHC. As it is often the case, the experiment finds more than theory predicts, and I highlight the discovery of the ``perfectly" liquid quark-gluon plasma at RHIC. I conclude with an outline of future opportunities, especially the search for a critical point in the QCD phase diagram.
\end{abstract}

\maketitle

%%%%%%%%%%%%%%%%%%%%%%%%%%%%%%%%%%%%%%%%%%%%%%%%%%%%%%%%%%%%%%%%%%%%%%%%%%%%%%%%%%%%%%%
\section{From Hagedorn to Quark-Gluon Plasma} 

%%%%%%%%%%%%%%%%%%%%%%%%%%%%%%%%%%%%%%%%%%%%%%%%%%%%%%%%%%%%%%%%%%%%%%%%%%%%%%%%%%%%%%%
\subsection*{Deconfinement of quarks and gluons} 

While successfully describing many features of multiparticle production at the energies accessible in the late 1960s, Hagedorn's statistical bootstrap model~\cite{Hagedorn:1965stB} with its exponentially growing mass spectrum of hadrons posed a quandary for cosmology~\cite{Huang:1970iqB}. The discovery of the cosmic microwave background in 1965 had provided unambiguous evidence for the hot Big-Bang model. By tracing back the cosmic evolution to very early times it was possible to conclude that the universe must have experienced temperatures in excess of 200 MeV at times less than 10 $\mu$s after the initial Big Bang. But what was the structure of the matter that filled the universe at such early times? What was its equation of state?

An exponential mass spectrum implied that the equation of state of hadronic matter has a singularity at the Hagedorn temperature, with empirical values in the range $150~{\rm MeV} < T_H < 200~{\rm MeV}$. Asking what the structure of matter at temperatures greater than $T_H$ is was meaningless in the statistical bootstrap model. The resolution of this quandary began with Collins and Perry's observation~\cite{Collins:1974kyB} in early 1975 that the asymptotic freedom of QCD implies that quarks are weakly interacting at short distances and therefore matter at very large quark densities should be composed of unconfined quarks. However, although they note that this argument should apply to matter in the early universe, their discussion is mostly focused on cold QCD matter.  

Later in the same year, Cabibbo and Parisi~\cite{Cabibbo:1975igB} proposed an interpretation to the singularity in the equation of state of Hagedorn's hadronic resonance gas as the point where strongly interacting matter changes from a gas of hadrons to a colored plasma of quarks and gluons. The Hagedorn temperature thus acquired the meaning of the critical temperature $T_c$ at which the composition of strongly interacting matter undergoes a discontinuous transition.\footnote{We now know that the exponentially growing mass spectrum of QCD is not related to a second order phase transition, as Cabibbo and Parisi surmised, but connected with the fact that QCD has an (approximate) string dual. In fact, lattice QCD has conclusively shown that the equation of state of QCD at zero or small net baryon density does not exhibit a singularity.} Quantitative predictions were impossible in the 1970s because of the lack of reliable mathematical or numerical techniques to solve QCD. 

%%%%%%%%%%%%%%%%%%%%%%%%%%%%%%%%%%%%%%%%%%%%%%%%%%%%%%%%%%%%%%%%%%%%%%%%%%%%%%%%%%%%%%%
\subsection*{Lattice QCD results} 

Starting in the early 1980s, Monte-Carlo simulations of the partition function of lattice QCD, first for the pure gauge theory and later for full QCD, made it possible to calculate the equation of state of strongly interacting matter {\em ab initio}. These calculations, which have recently converged to a definitive result~\cite{Borsanyi:2010cjB,Borsanyi:2013biaB,Bazavov:2014pvzB}, showed that matter composed of hadronic resonances is not separated from the quark-gluon plasma by a discontinuous phase transition in the absence of a baryon excess. However, a quasi-critical temperature $T_c \approx 155$ MeV can be defined as the temperature at which the chiral susceptibility -- the susceptibility associated with the scalar quark density $\langle\bar\psi\psi\rangle$ -- peaks.  The smooth cross-over is expected to turn into a first-order phase transition in the traditional sense of statistical physics for matter with a large baryon excess.

The lattice simulations showed that Hagedorn's model of a hadron resonance gas with an exponentially growing mass spectrum describes the equation of state of QCD matter and many other observables very well for temperatures below $T_c$. The precision of the lattice QCD simulations is now good enough to distinguish between the equation of state of a hadron gas made up of the resonances tabulated in the Particle Data Book or that of a Hagedorn resonance gas. The numerical results point to a continuation of the exponential growth of the hadron mass spectrum beyond the reach of direct detection of resonances and thus support Hagedorn's hadronic bootstrap model~\cite{Majumder:2010ikB}. 

The implied existence of many unknown hadron resonances may also be present in the strange baryon sector~\cite{Bazavov:2014xyaB}. Above $T_c$ the density of states grows much less rapidly and eventually approaches that of a perturbatively interacting quark-gluon plasma composed of massive quasiparticles, confirming the notion that the Hagedorn temperature signals the transition from a hadron resonance gas to a new state of matter.

%%%%%%%%%%%%%%%%%%%%%%%%%%%%%%%%%%%%%%%%%%%%%%%%%%%%%%%%%%%%%%%%%%%%%%%%%%%%%%%%%%%%%%%

\subsection*{Hot nuclear matter} 

The next critical step was the realization, arising most prominently from discussions in the CERN Theory Division\footnote{What distinguished these discussions from other theoretical speculation in the mid-1970s was that the focus was on {\em thermal} properties of strongly interacting matter, rather than properties of compressed baryonic matter (see e.g. Lee~\cite{BearMountain:1974B,Lee:1974knB}.)}, that temperatures in the range of $T_c$ and even beyond could be created in the laboratory by colliding heavy atomic nuclei at sufficiently high energies. 

The experimental study of relativistic heavy ion collisions with stationary targets had commenced at the Bevalac in the mid-1970s, but the energies available there were recognized to be insufficient to reach $T_c$. The CERN SPS could provide much higher energies, and back-of-the-envelope calculations suggested that temperatures near and above $T_c$ would be reached if the nuclear matter in the colliding nuclei thermalized rapidly. Hagedorn and Rafelski extended the statistical bootstrap model to matter with a baryon excess and found that under certain assumptions the equation of state exhibited a first-order phase transition~\cite{Hagedorn:1980kbB}. 

I had the good fortune of meeting Hagedorn during several visits with Johann at CERN during this formative period in the late 1970s. My conversations with them inspired my own interest in hot QCD and soon thereafter resulted in our joint work on the thermal properties of the QCD vacuum~\cite{Muller:1980kfB} and on particle production with exact symmetry in proton-antiproton annihilation~\cite{Muller:1982gdB}. What impressed me most on these occasions was Hagedorn's willingness to share his thoughts with a young scientist without imposing on him. One puzzling aspect of the experimental observation of thermal particle emission that is still occupying theorists today -- how a large fraction of the kinetic energy carried by the incident particles could be thermalized within a time of order 1 fm/c -- led to my interest in the chaotic properties of non-abelian gauge theories. I vividly recall Hagedorn's excitement after he listened to my talk about our numerical studies of dynamical chaos of the Yang-Mills field at the workshop in Divonne~\cite{BMDivonne}.

%%%%%%%%%%%%%%%%%%%%%%%%%%%%%%%%%%%%%%%%%%%%%%%%%%%%%%%%%%%%%%%%%%%%%%%%%%%%%%%%%%%%%%%

\section{Path to Discovery of the QGP} 

%%%%%%%%%%%%%%%%%%%%%%%%%%%%%%%%%%%%%%%%%%%%%%%%%%%%%%%%%%%%%%%%%%%%%%%%%%%%%%%%%%%%%%%

\subsection*{QGP observables} 

The biggest challenge on the way to discovery was finding signatures that could provide evidence that nuclear matter had made the transition to a quark-gluon plasma for a brief period during the collision. One either had to look at penetrating probes, such as photons and lepton pairs~\cite{Feinberg:1976uaB}, that could escape from the hot fireball, or at probes that retained their identity under the action of the strong interactions in the final state, such as quark flavor.  

Shuryak took the matter further by evoking quark and gluon degrees of freedom in \pp reactions and focusing on electromagnetic probes and charm quarks as signatures for the formation of a thermal QCD plasma~\cite{Shuryak:1977utB,Shuryak:1978ijB}. Rafelski, in collaboration with Hagedorn, Danos, and myself, focused on strange quarks whose mass is sufficiently low for them to be produced thermally in the quark-gluon plasma~\cite{Rafelski:1980rkB}. 

The strangeness argument was not simply that strange quarks and antiquarks would be produced abundantly at temperatures above $T_c$, but that baryons containing multiple strange quarks would be produced copiously and in chemical equilibrium when the quark-gluon plasma hadronizes by recombination of the deconfined quarks into hadrons. A calculation of thermal strange quark pair production in the quark-gluon plasma~\cite{Rafelski:1982puB} confirmed that flavor equilibrium could, indeed, be reached on the time scales of a relativistic heavy ion collision and showed that thermal gluons played a crucial role in the flavor equilibration process.

Following on the recognition of the abundant strangeness in quark-gluon plasma, Johann and I embarked on the task of developing a bulk hadronization model that would enable us to make quantitative predictions for the strange antibaryon signature of the quark-gluon plasma. Our effort grew over two years, in collaboration with Peter Koch, into a Physics Reports article~\cite{Koch:1986udB}. Among the highlights of this work is the development of the recombination and fragmentation-recombination models of quark-gluon hadronization that in slightly modified form remain in use today~\cite{Fries:2011wz}. We enforced conservation laws, assured increase of entropy, and quantified the production of strange (anti-)baryons with their strangeness content. These developments set clear experimental goals for the forthcoming SPS strangeness experiments  which are further discussed below and in the contribution of Emanuele Quercigh.

%%%%%%%%%%%%%%%%%%%%%%%%%%%%%%%%%%%%%%%%%%%%%%%%%%%%%%%%%%%%%%%%%%%%%%%%%%%%%%%%%%%%%%%

\subsection*{SPS results} 

The heavy ion experiments at the SPS, which commenced in 1986/87, impressively confirmed these ideas. The chemical composition of the hadrons emitted from the collisions can be well described by a chemical near equilibrium gas at a temperature close to $T_c$ and a baryon chemical potential that varies strongly with the collision energy~\cite{Becattini:2003wpB}. The strong enhancement and full chemical equilibration of baryons and anti-baryons containing multiple strange quarks~\cite{Andersen:1999ymB,Antinori:1999rwB} could only be explained if hadrons containing valence quarks of all three light flavors were ``born'' into thermal abundances~\cite{Letessier:1998nq,Heinz:1999kb,Stock:1999hm}. 

However, the SPS data did not provide other corroborating evidence for the existence of a thermal phase of matter at temperatures above $T_c$ from which these hadrons formed by statistical emission. The (unpublished) CERN announcement of a new state of matter~\cite{Heinz:2000bkB} in 2000 was thus greeted with skepticism by many physicists.
Experiments with heavy ion collisions at much higher energies were needed to resolve this issue. 

%%%%%%%%%%%%%%%%%%%%%%%%%%%%%%%%%%%%%%%%%%%%%%%%%%%%%%%%%%%%%%%%%%%%%%%%%%%%%%%%%%%%%%%

\subsection*{Experiments at RHIC} 

Commencing at RHIC in year 2000, these experiments allowed to access a new kinematic domain, in which the produced matter is imprinted from the start with a nearly boost invariant longitudinal flow profile. An analytical solution of relativistic hydrodynamics for this initial condition had been found by Bjorken~\cite{Bjorken:1982qrB}, and it provided the basis for a systematic investigation of the collective properties of the matter formed in the nuclear collisions~\cite{Arsene:2004faB,Adcox:2004mhB,Back:2004jeB,Adams:2005dqB}. The fact that the transverse geometric profile of the reaction zone and the initial energy density fluctuations from event to event could be correlated with the patterns observed in the collective flow of the emitted hadrons made it possible to pin down the transport properties of the expanding matter, which was shown to have an extraordinarily low shear viscosity, relative to its entropy density~\cite{Teaney:2003kpB,Romatschke:2007mq,Song:2010mg}. The matter was thus shown to be a liquid at temperatures well above $T_c$. 

A detailed study of the subtle variations of the flow profile between different hadron species revealed that these variations disappeared when all hadrons were assumed to be formed by recombination of deconfined, collectively flowing quarks when the matter cooled below $T_c$~\cite{Fries:2003vbB}. Together, these observations provided strong evidence for the notion that the matter formed in nuclear collisions at RHIC is, indeed, a plasma of deconfined quarks and gluons, which behaves as a nearly inviscid liquid and decays by the emission of hadrons in chemical and thermal equilibrium. Because the matter is already expanding very rapidly when the transition to a hadron gas occurs, many observables are nearly unaffected by final-state interactions among hadrons. The low viscosity of the liquid quark-gluon plasma implies that the interactions among quarks and gluons contained in it are strong. Other observations, such as the strong suppression of high-momentum hadrons and of charmonium, support this conclusion (for early reviews, see:~\cite{Gyulassy:2004vg,Muller:2006eeB}).

%%%%%%%%%%%%%%%%%%%%%%%%%%%%%%%%%%%%%%%%%%%%%%%%%%%%%%%%%%%%%%%%%%%%%%%%%%%%%%%%%%%%%%%

\subsection*{Experiments at LHC} 

Experiments at even higher energies at the LHC have impressively confirmed the nature of QCD matter above $T_c$ as a strongly coupled, liquid quark-gluon plasma~\cite{Muller:2012zqB}. A careful analysis of the LHC data revealed that the average strong coupling at the higher energy density reached at LHC is slightly weaker than at RHIC~\cite{Betz:2012qqB}, in accordance with the running of $\alpha_s$ with temperature. The reduced coupling is also reflected in a somewhat larger shear viscosity-to-entropy density ratio~\cite{Gale:2012rqB}. 

In addition to consolidating the insights gained at RHIC, the much higher energy available at LHC permit more detailed studies of the event-by-event fluctuations of the collective flow pattern, which reflect the quantum fluctuations of the initial energy density distribution. Enabled by the design of the LHC detectors, the higher energy also allows for precise studies of the phenomenon of jet quenching that was first discovered at RHIC. And finally, the large yield of primordially produced charm quarks at LHC results in abundant late-stage recombination of charm-anticharm quark pairs into charmonium, providing additional evidence for the deconfinement of quarks in the QCD plasma phase.

%%%%%%%%%%%%%%%%%%%%%%%%%%%%%%%%%%%%%%%%%%%%%%%%%%%%%%%%%%%%%%%%%%%%%%%%%%%%%%%%%%%%%%%

\subsection*{Beam energy scan at RHIC} 

How far down in beam energy does the phenomenology discovered and established at RHIC persist? Where is the threshold below which no quark-gluon plasma is formed?  Did the SPS experiments produce a quark-gluon plasma? In order to address these open questions, RHIC has recently collided heavy ions at lower energies, down to $\sqrt{s_{\rm NN}} = 7.7$ GeV. An extensive analysis of the data gathered in this beam energy scan is now available~\cite{Sorensen:2013jkaB,Sahoo:2014bqaB}. It shows that the matter produced in collisions down to the top SPS energy, $\sqrt{s_{\rm NN}} = 19.6$ GeV, exhibits some of the same characteristics as that produced at the top RHIC energy, $\sqrt{s_{\rm NN}} = 200$ GeV. 

However, there are noticeable differences. Matter produced at the lower beam energies contains a larger excess of baryons resulting in a different chemical composition of the emitted hadrons; energetic hadrons are no longer suppressed at lower energies; and no direct photon signal has been observed. Thus it is quite likely that the CERN experiments succeeded in breaking through the thermal barrier of the Hagedorn temperature, but it is still unclear what kind of baryon-rich matter they produced and whether it exhibited collective behavior at the parton level. Theoretical models that can more reliably describe nuclear reactions at these lower energies will be needed to finally address this issue.

%%%%%%%%%%%%%%%%%%%%%%%%%%%%%%%%%%%%%%%%%%%%%%%%%%%%%%%%%%%%%%%%%%%%%%%%%%%%%%%%%%%%%%%

\subsection*{Next steps} 

Where do we go from here? Two major questions remain to be answered: (1) Is there a critical point in the phase diagram of QCD matter where the cross-over from hadron resonance gas to the quark-gluon plasma turns into a true phase transition, and where is it located in $T$ and $\mu$? (2) What are the effective constituents of the liquid quark-gluon plasma?

The first question will be addressed in a second, high statistics beam energy scan that is planned to be carried out in 2018--19 at RHIC after a luminosity upgrade of the collider at low beam energies. Physicists will then look for telltale signs of a phase transition, including critical fluctuations in baryon number or large event-by-event fluctuations caused by spinodal decomposition of the matter at the phase boundary. A recent discussion of the theoretical and experimental challenges of locating the QCD critical point can be found in~\cite{Gavai:2014elaB}.

Addressing the second question requires probes that are sensitive to the structure of the quark-gluon plasma at shorter than thermal length scales. Two such probes are heavy quarks and jets. The experiments at LHC and now at RHIC are equipped with powerful vertex detectors that can identify hadrons containing heavy quarks. They will study the transport of $c$ and $b$ quarks in the plasma in great detail and hopefully detect clues to its internal structure. Jets explore multiple length scales as they develop inside the matter after the initial hard scattering event. Extensive jet measurement programs, which are already underway at the LHC, are planned for RHIC in the decade ahead~\cite{Sickles:2013ilaB}.

%%%%%%%%%%%%%%%%%%%%%%%%%%%%%%%%%%%%%%%%%%%%%%%%%%%%%%%%%%%%%%%%%%%%%%%%%%%%%%%%%%%%%%%

\section{Outlook and Conclusions}

Our understanding of the structure and properties of hadronic matter at high energy density has made tremendous progress since the days when the question first arose in full urgency in the late 1960s, and remarkable discoveries have been made along the way. We have established that Hagedorn's gas of hadron resonances turns into a liquid quark-gluon plasma when heated above 155 MeV, quite an extraordinary phenomenon in itself. We have discovered a liquid that comes very close to the quantum bound on the shear viscosity imposed by unitarity. And we have learned that the statistical and collective properties of the flowing quark-gluon plasma get imprinted onto the emitted hadrons in a characteristic way that makes it possible to experimentally determine the thermal and chemical properties of the QCD phase boundary. Rolf Hagedorn would surely be satisfied to witness that the questions he helped pose fifty years ago have proved to be so extraordinarily fertile.

%%%%%%%%%%%%%%%%%%%%%%%%%%%%%%%%%%%%%%%%%%%%%%%%%%%%%%%%%%%%%%%%%%%%%%%%%%%%%%%%%%%%%%%

{\em Acknowledgement:} I thank W.~A.~Zajc, J.~Rafelski, and U.~Heinz for constructive comments. This work was supported by Grant no. DE-FG02-05ER41367 from the U.~S.~Department of Energy.

%%%%%%%%%%%%%%%%%%%%%%%%%%%%%%%%%%%%%%%%%%%%%%%%%%%%%%%%%%%%%%%%%%%%%%%%%%%%%%%%%%%%%%%

%%%%%%%%%%%%%%%%%%%%%%%%%%%%%%%%%%%%%%%%%


\begin{thebibliography}{99}

\bibitem{Hagedorn:1965stB} 
  R.~Hagedorn:
  ``Statistical thermodynamics of strong interactions at high-energies,''
  Nuovo Cim.\ Suppl.\  {\bf 3}, 147 (1965)
  
\bibitem{Huang:1970iqB} 
  K.~Huang and S.~Weinberg:
  ``Ultimate temperature and the early universe,''
  Phys.\ Rev.\ Lett.\  {\bf 25}, 895 (1970)

\bibitem{Collins:1974kyB} 
  J.~C.~Collins and M.~J.~Perry:
  ``Superdense Matter: Neutrons Or Asymptotically Free Quarks?,''
  Phys.\ Rev.\ Lett.\  {\bf 34}, 1353 (1975)
  
\bibitem{Cabibbo:1975igB} 
  N.~Cabibbo and G.~Parisi:
  ``Exponential Hadronic Spectrum and Quark Liberation,''
  Phys.\ Lett.\ B {\bf 59}, 67 (1975)

\bibitem{Borsanyi:2010cjB} 
  S.~Borsanyi, G.~Endrodi, Z.~Fodor, A.~Jakovac, S.~D.~Katz, S.~Krieg, C.~Ratti and K.~K.~Szabo:
  ``The QCD equation of state with dynamical quarks,''
  JHEP {\bf 1011}, 077 (2010)
%  [arXiv:1007.2580 [hep-lat]].
  
\bibitem{Borsanyi:2013biaB} 
  S.~Borsanyi, Z.~Fodor, C.~Hoelbling, S.~D.~Katz, S.~Krieg and K.~K.~Szabo:
  ``Full result for the QCD equation of state with 2+1 flavors,''
  Phys.\ Lett.\ B {\bf 730}, 99 (2014)
%  [arXiv:1309.5258 [hep-lat]].

\bibitem{Bazavov:2014pvzB} 
  A.~Bazavov {\it et al.}  [HotQCD Collaboration]:
  ``Equation of state in ( 2+1 )-flavor QCD,''
  Phys.\ Rev.\ D {\bf 90}, no. 9, 094503 (2014)
%  [arXiv:1407.6387 [hep-lat]].
  
\bibitem{Majumder:2010ikB} 
  A.~Majumder and B.~M\"uller:
  ``Hadron Mass Spectrum from Lattice QCD,''
  Phys.\ Rev.\ Lett.\  {\bf 105}, 252002 (2010)
%  [arXiv:1008.1747 [hep-ph]].
  
\bibitem{Bazavov:2014xyaB} 
  A.~Bazavov, H.-T.~Ding, P.~Hegde, O.~Kaczmarek, F.~Karsch, E.~Laermann, Y.~Maezawa and S.~Mukherjee {\it et al.}:
  ``Additional Strange Hadrons from QCD Thermodynamics and Strangeness Freezeout in Heavy Ion Collisions,''
  Phys.\ Rev.\ Lett.\  {\bf 113}, no. 7, 072001 (2014)
%  [arXiv:1404.6511 [hep-lat]].

\bibitem{BearMountain:1974B}
  {\em Report of the workshop on BeV/nucleon collisions of heavy ions -- how and why}, 
  Bear Mountain, New York, Nov. 29-Dec. 1, 1974 (BNL-AUI, 1975).
  
\bibitem{Lee:1974knB} 
  T.~D.~Lee:
  ``Abnormal Nuclear States and Vacuum Excitations,''
  Rev.\ Mod.\ Phys.\  {\bf 47}, 267 (1975)
  
\bibitem{Hagedorn:1980kbB} 
  R.~Hagedorn and J.~Rafelski:
  ``Hot Hadronic Matter and Nuclear Collisions,''
  Phys.\ Lett.\ B {\bf 97}, 136 (1980)

\bibitem{Muller:1980kfB} 
  B.~M\"uller and J.~Rafelski:
  ``Temperature Dependence of the Bag Constant and the Effective Lagrangian for Gauge Fields at Finite Temperatures,''
  Phys.\ Lett.\ B {\bf 101}, 111 (1981) 

\bibitem{Muller:1982gdB} 
  B.~M\"uller and J.~Rafelski:
  ``Role of Internal Symmetry in $p \bar{p}$ Annihilation,''
  Phys.\ Lett.\ B {\bf 116}, 274 (1982) 
  
\bibitem{BMDivonne}
  B.~M\"uller: ``Colored Chaos,''
  in: {\it Hot Hadronic Matter: Theory and Experiment},  
  Divonne 1994, J. Letessier, H.H. Gutbrod and J. Rafelski, eds.   
  NATO-ASI  {\bf 346}, pp. 171-180, (Plenum Press, New York  1995)

\bibitem{Feinberg:1976uaB} 
  E.~L.~Feinberg:
 ``Direct Production of Photons and Dileptons in Thermodynamical Models of Multiple Hadron Production,''
  Nuovo Cim.\ A {\bf 34}, 391 (1976)

\bibitem{Shuryak:1977utB} 
  E.~V.~Shuryak:
  ``Theory of Hadronic Plasma,''
  Sov.\ Phys.\ JETP {\bf 47}, 212 (1978)
  [Zh.\ Eksp.\ Teor.\ Fiz.\  {\bf 74}, 408 (1978)]

\bibitem{Shuryak:1978ijB} 
  E.~V.~Shuryak:
  ``Quark-Gluon Plasma and Hadronic Production of Leptons, Photons and Psions,''
  Phys.\ Lett.\ B {\bf 78}, 150 (1978)
%  [Sov.\ J.\ Nucl.\ Phys.\  {\bf 28}, 408 (1978); Yad.\ Fiz.\  {\bf 28}, 796 (1978)]

\bibitem{Rafelski:1980rkB} 
  J.~Rafelski and R.~Hagedorn:
 ``From Hadron Gas to Quark Matter. 2.,'' pp. 253-272
  In ``Bielefeld 1980, Statistical Mechanics Of Quarks and Hadrons'' H. Satz, edt. (Amsterdam 1980) 

\bibitem{Rafelski:1982puB} 
  J.~Rafelski and B.~M\"uller:
 ``Strangeness Production in the Quark-Gluon Plasma,''
  Phys.\ Rev.\ Lett.\  {\bf 48}, 1066 (1982)
  [Erratum-ibid.\  {\bf 56}, 2334 (1986)]

\bibitem{Koch:1986udB} 
  P.~Koch, B.~M\"uller and J.~Rafelski,
  ``Strangeness in Relativistic Heavy Ion Collisions,''
  Phys.\ Rept.\  {\bf 142}, 167 (1986).

\bibitem{Fries:2011wz} 
  R.~J.~Fries:
  ``Quark Recombination in Heavy Ion Collisions,''
  PoS CERP {\bf 2010}, 008 (2010) 
%  [arXiv:1102.5723 [nucl-th]] 

\bibitem{Becattini:2003wpB} 
  F.~Becattini, M.~Gazdzicki, A.~Keranen, J.~Manninen and R.~Stock:
 ``Chemical equilibrium in nucleus nucleus collisions at relativistic energies,''
  Phys.\ Rev.\ C {\bf 69}, 024905 (2004)
%  [hep-ph/0310049].
  
\bibitem{Andersen:1999ymB} 
  E.~Andersen {\it et al.}  [WA97 Collaboration]:
  ``Strangeness enhancement at mid-rapidity in Pb Pb collisions at 158-A-GeV/c,''
  Phys.\ Lett.\ B {\bf 449}, 401 (1999)
  
\bibitem{Antinori:1999rwB} 
  F.~Antinori {\it et al.}  [WA97 Collaboration]:
  ``Strangeness enhancement at mid-rapidity in Pb Pb collisions at 158-A-GeV/c: A comparison with VENUS and RQMD models,''
  Eur.\ Phys.\ J.\ C {\bf 11}, 79 (1999) 
  
\bibitem{Letessier:1998nq} 
  J.~Letessier and J.~Rafelski:
  ``Evidence for QGP in Pb Pb 158-A-Gev collisions from strange particle abundances and the Coulomb effect,''
  Acta Phys.\ Polon.\ B {\bf 30}, 153 (1999)
%  [hep-ph/9807346].

\bibitem{Stock:1999hm} 
  R.~Stock:
  ``The parton to hadron phase transition observed in Pb + Pb collisions at 158-GeV per nucleon,''
  Phys.\ Lett.\ B {\bf 456}, 277 (1999)
%  [hep-ph/9905247].
  
\bibitem{Heinz:1999kb} 
  U.~W.~Heinz:
  ``Primordial hadrosynthesis in the Little Bang,''
  Nucl.\ Phys.\ A {\bf 661}, 140 (1999)
%  [nucl-th/9907060].

\bibitem{Heinz:2000bkB} 
  U.~W.~Heinz and M.~Jacob:
 ``Evidence for a new state of matter: An Assessment of the results from the CERN lead beam program,'' (CERN February 2000 press release)
%  nucl-th/0002042.
  
\bibitem{Bjorken:1982qrB} 
  J.~D.~Bjorken:
  ``Highly Relativistic Nucleus-Nucleus Collisions: The Central Rapidity Region,''
  Phys.\ Rev.\ D {\bf 27}, 140 (1983)
    
\bibitem{Arsene:2004faB} 
  I.~Arsene {\it et al.}  [BRAHMS Collaboration]:
  ``Quark gluon plasma and color glass condensate at RHIC? The Perspective from the BRAHMS experiment,''
  Nucl.\ Phys.\ A {\bf 757}, 1 (2005)
%  [nucl-ex/0410020].
  
\bibitem{Adcox:2004mhB} 
  K.~Adcox {\it et al.}  [PHENIX Collaboration]:
 ``Formation of dense partonic matter in relativistic nucleus-nucleus collisions at RHIC: Experimental evaluation by the PHENIX collaboration,''
  Nucl.\ Phys.\ A {\bf 757}, 184 (2005)
%  [nucl-ex/0410003].
  
\bibitem{Back:2004jeB} 
  B.~B.~Back {\it et al.} [PHOBOS Collaboration]:
  ``The PHOBOS perspective on discoveries at RHIC,''
  Nucl.\ Phys.\ A {\bf 757}, 28 (2005)
%  [nucl-ex/0410022].
  
  \bibitem{Adams:2005dqB} 
  J.~Adams {\it et al.}  [STAR Collaboration]:
  ``Experimental and theoretical challenges in the search for the quark gluon plasma: The STAR Collaboration's critical assessment of the evidence from RHIC collisions,''
  Nucl.\ Phys.\ A {\bf 757}, 102 (2005)
%  [nucl-ex/0501009].

\bibitem{Teaney:2003kpB} 
  D.~Teaney:
 ``The Effects of viscosity on spectra, elliptic flow, and HBT radii,''
  Phys.\ Rev.\ C {\bf 68}, 034913 (2003)
%  [nucl-th/0301099].

\bibitem{Romatschke:2007mq} 
  P.~Romatschke and U.~Romatschke:
  ``Viscosity Information from Relativistic Nuclear Collisions: How Perfect is the Fluid Observed at RHIC?,''
  Phys.\ Rev.\ Lett.\  {\bf 99}, 172301 (2007)
%  [arXiv:0706.1522 [nucl-th]].

\bibitem{Song:2010mg} 
  H.~Song, S.~A.~Bass, U.~Heinz, T.~Hirano and C.~Shen:
  ``200 A GeV Au+Au collisions serve a nearly perfect quark-gluon liquid,''
  Phys.\ Rev.\ Lett.\  {\bf 106}, 192301 (2011)
  [Erratum-ibid.\  {\bf 109}, 139904 (2012)]
%  [arXiv:1011.2783 [nucl-th]].

\bibitem{Fries:2003vbB} 
  R.~J.~Fries, B.~M\"uller, C.~Nonaka and S.~A.~Bass:
  ``Hadronization in heavy ion collisions: Recombination and fragmentation of partons,''
  Phys.\ Rev.\ Lett.\  {\bf 90}, 202303 (2003) and 
%  [nucl-th/0301087];
%\bibitem{Fries:2003kq}     
  ``Hadron production in heavy ion collisions: Fragmentation and recombination from a dense parton phase,''
  Phys.\ Rev.\ C {\bf 68}, 044902 (2003)
%  [nucl-th/0306027].

\bibitem{Gyulassy:2004vg} 
  M.~Gyulassy,
  ``The QGP discovered at RHIC,''
  nucl-th/0403032.
  
\bibitem{Muller:2006eeB} 
  B.~M\"uller and J.~L.~Nagle:
 ``Results from the relativistic heavy ion collider,''
  Ann.\ Rev.\ Nucl.\ Part.\ Sci.\  {\bf 56}, 93 (2006)
%  [nucl-th/0602029].  

\bibitem{Muller:2012zqB} 
  B.~M\"uller, J.~Schukraft and B.~Wyslouch,
 ``First Results from Pb+Pb collisions at the LHC,''
  Ann.\ Rev.\ Nucl.\ Part.\ Sci.\  {\bf 62}, 361 (2012)
%  [arXiv:1202.3233 [hep-ex]].
 
 \bibitem{Betz:2012qqB} 
  B.~Betz and M.~Gyulassy:
  ``Examining a reduced jet-medium coupling in Pb+Pb collisions at the Large Hadron Collider,''
  Phys.\ Rev.\ C {\bf 86}, 024903 (2012)
%  [arXiv:1201.0281 [nucl-th]].

\bibitem{Gale:2012rqB} 
  C.~Gale, S.~Jeon, B.~Schenke, P.~Tribedy and R.~Venugopalan:
  ``Event-by-event anisotropic flow in heavy-ion collisions from combined Yang-Mills and viscous fluid dynamics,''
  Phys.\ Rev.\ Lett.\  {\bf 110}, 012302 (2013)
%  [arXiv:1209.6330 [nucl-th]].

\bibitem{Sorensen:2013jkaB} 
  P.~Sorensen:
  ``Beam Energy Scan Results from RHIC,''
  J.\ Phys.\ Conf.\ Ser.\  {\bf 446}, 012015 (2013).

\bibitem{Sahoo:2014bqaB} 
  N.~R.~Sahoo [STAR Collaboration]:
  ``Recent results on event-by-event fluctuations from the RHIC Beam Energy Scan program in the STAR experiment,''
  J.\ Phys.\ Conf.\ Ser.\  {\bf 535}, 012007 (2014)
%  [arXiv:1407.1554 [nucl-ex]].

\bibitem{Adare:2014qvsB} 
  A.~Adare {\it et al.}  [PHENIX Collaboration]:
  ``Beam-energy and system-size dependence of the space-time extent of the pion emission source produced in heavy ion collisions,''
  arXiv:1410.2559 [nucl-ex].
  
  \bibitem{Gavai:2014elaB} 
  R.~V.~Gavai:
  ``QCD Critical Point: The Race is On,''
  arXiv:1404.6615 [hep-ph] 

\bibitem{Sickles:2013ilaB} 
  A.~M.~Sickles {\it et al.}  [PHENIX Collaboration]:
  ``The Physics of sPHENIX,''
  J.\ Phys.\ Conf.\ Ser.\  {\bf 446}, 012050 (2013) 

\end{thebibliography}
\end{document}